\documentclass[article, twocolumn,
   aps, pra,
  amsmath,amssymb,
  longbibliography,
  ]{revtex4-1}
\usepackage{graphicx,color}
\usepackage{amsmath}
\usepackage{natbib}
\usepackage{epsfig}
\begin{document}

\title{Heat capacity of dense liquids: A link between two-phase model and melting temperature scaling}

\author{S. A. Khrapak\email{Sergey.Khrapak@gmx.de} 
and A. G. Khrapak
}
\affiliation{Joint Institute for High Temperatures, Russian Academy of Sciences, 125412 Moscow, Russia}

\begin{abstract}
Generalized Rosenfeld-Tarazona scaling predicts the power-law dependence of the excess heat capacity of simple liquids on temperature. The two-phase model treats a liquid as a superposition of gas- and solid-like components whose relative abundance is quantified by a liquid rigidity parameter. We demonstrate here that the generalized Rosenfeld-Tarazona scaling emerges naturally in the two-phase model from the scale invariance of the liquid rigidity parameter.    
\end{abstract}

\date{\today}

\maketitle

Heat capacity is an important property of matter, which is much better understood in solids and gases as compared to liquids~\cite{TrachenkoBook}. According to statistical physics~\cite{LandauStatPhys}, the isochoric (constant volume) heat capacity of a monatomic ideal gas is determined by $3N$ translational degrees of freedom and is $C_{\rm V}=\tfrac{3}{2}Nk_{\rm B}$, where $N$ is the number of particles and $k_{\rm B}$ is the Boltzmann constant. In an ideal harmonic classical crystal, the heat capacity is determined by $3N$ translational degrees of freedom and $3N$ vibrational degrees of freedom, and hence $C_{\rm V}=3Nk_{\rm B}$. This result is known as the Dulong-Petit law. A notable exemption is a system of hard spheres, where, due to the absence of potential energy of interaction between the spheres, the heat capacity is always equal to that of an ideal gas. 

How exactly the transition between solid- and gas-like limits occurs in the liquid regime remains an important open question, even for simple monatomic liquids far from the critical point~\cite{TrachenkoBook}. Several different theoretical approaches have been discussed in the literature. One of them is the melting temperature scaling of the excess thermal energy, also known as the Rosenfeld-Tarazona (RT) scaling~\cite{RosenfeldMolPhys1998,RosenfeldPRE2000}.
Based on this scaling, a decay of the excess isochoric
heat capacity $C_{\rm V}^{\rm ex}\propto T^{-2/5}$ is predicted. This appears a good approximation, in particular for liquids with strong correlations
between equilibrium fluctuations of virial and potential energy~\cite{IngebrigtsenJCP2013}, known as ``Roskilde-simple liquids ''~\cite{IngebrigtsenPRX2012,DyreJPCB2014}.  
The physical mechanisms behind this particular power-like decay have remained elusive. Moreover, empirical evidence suggests that different exponents are sometimes more appropriate and that the generalized RT scaling $C_{\rm V}^{\rm ex}\propto T^{-\alpha}$ works better~\cite{KhrapakPRE09_2024,KhrapakPOF11_2024}.  

Another popular recent approach is the phonon theory of liquid thermodynamics~\cite{TrachenkoPRB2008,
BolmatovSciRep2012,BolmatovAP2015,TrachenkoRPP2015,BolmatovJPCL2022,LiuPRB2025}. The theory stems from the ideas discussed by Frenkel~\cite{FrenkelBook} that dense liquids can be approached from a solid-state perspective (see also Ref.~\cite{KhrapakPhysRep2024} for a recent review about the vibrational paradigm of liquid dynamics). Combining a similarity between the high-frequency elastic properties of liquids and solids with the Debye vibrational density of states (or approximations that are more appropriate for the liquid state) allows the calculation of the heat capacities of various liquids, which in some cases show reasonable agreement with the experimental results~\cite{TrachenkoPRB2008,
BolmatovSciRep2012,BolmatovAP2015,TrachenkoRPP2015,BolmatovJPCL2022,LiuPRB2025}. 
It should be mentioned that a careful analysis of the collective excitation spectra is essential in this approach. For some relevant recent developments, see Refs.~\cite{KhrapakJCP2019,KryuchkovSciRep2019,KryuchkovJCPL2019,YakovlevJCPL2020,KhrapakPRE09_2020}. The properties of dispersion of collective excitations in liquids and in particular of the transverse mode have a strong effect on the heat capacity~\cite{KryuchkovPRL2020}. Transverse collective excitations and related limitations of the phonon theory of liquid thermodynamics have recently been discussed~\cite{BrykPRE2025}. 

The two-phase model represents a liquid as a superposition of gas-like and solid-like states~\cite{LinJCP2003,PascalPCCP2011,MoonPRR2024,KhrapakPRE2025}. To be concrete, let us assume that the relative abundance of solid states is $x$. This implies that the relative abundance of gas-like states is $1-x$. Following a simple picture, presented in Ref.~\cite{MoonPRR2024} the thermal energy of a liquid can be written as
\begin{equation}\label{energy}
E = x(3Nk_{\rm B}T)+(1-x)\left(\frac{3}{2}Nk_{\rm B}T\right),
\end{equation}
where $T$ is the temperature. The isochoric heat capacity is then
\begin{equation}\label{CV}
C_{\rm V}=\frac{dE}{dT}=\frac{3}{2}Nk_{\rm B}+\frac{3}{2}xNk_{\rm B}+\frac{3}{2}Nk_{\rm B}T\frac{dx}{dT}.
\end{equation}
In reduced (dimensionless) units we get
\begin{equation}\label{c_v}
c_{\rm v}=\frac{C_{\rm V}}{Nk_{\rm B}}=\frac{3}{2}\left(1+x+\frac{dx}{d\ln T}\right).    
\end{equation}
The remaining step is to appropriately define the parameter $x$. Different procedures have been discussed in the literature. For example, Lin {\it et al}.~\cite{LinJCP2003} suggested decomposing the vibrational density of states (VDoS) extracted from the numerically simulated velocity autocorrelation function into solid-like (nondiffusive) and gas-like (diffusive) components. Their ``fluidity parameter'' $f=1-x$ can then be obtained as the ratio of the integral over diffusive modes and the integral over the full VDoS. In a recent paper Moon {\it et al}.~\cite{MoonPRR2024} have used instantaneous normal modes analysis to characterize various phases. From instantaneous structures, the normal mode spectra of liquids is reconstructed, which naturally contains imaginary modes (those with $\omega^2<0$) and real modes (those with $\omega^2>0$). The sum of imaginary modes $N_i$ and real modes $N_r$ gives the total number of normal modes $N_i+N_r=N_{\rm total}=3N$. To describe the transition from solid to gas, they then propose using a phenomenological ``instability factor'' (IF) defined as ${\rm IF}=N_i/N_r=1-x$ (actually, two slightly different definitions are proposed, yielding similar results; the details are not essential in the present context).       

The two approaches discussed above are relatively resource-consuming and, to some extent, arbitrary. Here we introduce an extremely simple practical method based on very general arguments. Importantly, our general approach will lead us to the generalized RT scaling, pointing out its origin.  

\begin{figure}
\includegraphics[width=8cm]{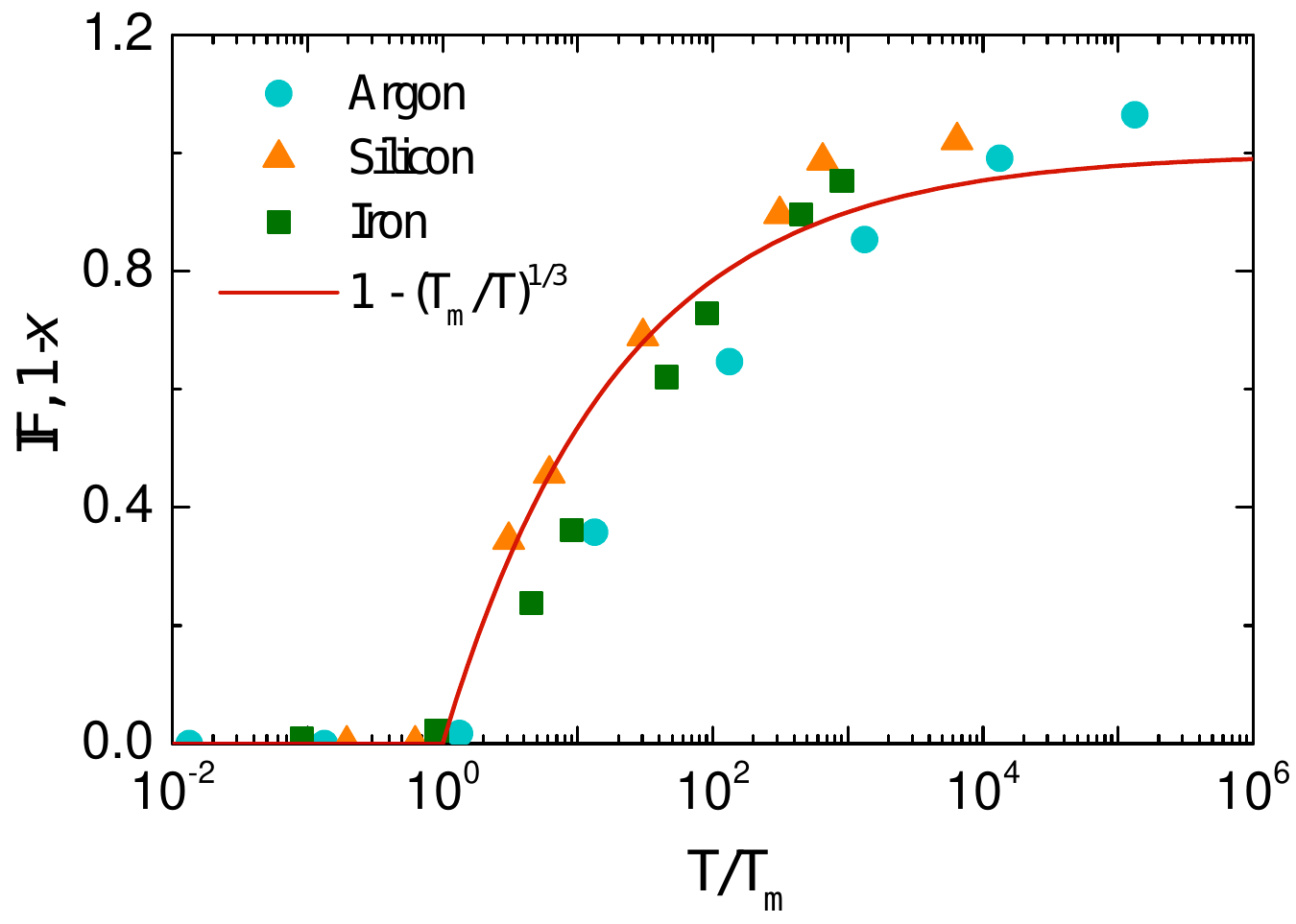}
\caption{(Color online) Relative abundance of the gas-like component versus the reduced temperature $T/T_{\rm m}$. Symbols correspond to the instability factor IF, calculated in Ref.~\cite{MoonPRR2024} for argon, silicon, and iron (data shown in Fig. 8(a) with melting temperatures from Fig. 4 of Ref.~\cite{MoonPRR2024}). The solid line corresponds to the liquid rigidity parameter proposed here, Eq.~(\ref{rigidity}) with $\alpha=1/3$.}
\label{Fig1}
\end{figure}

The relative abundance of the solid component $x$, which we find appropriate to call the liquid ``rigidity parameter'' should satisfy the following two limiting conditions: (i) in the high temperature limit the behavior of the ideal gas is recovered (we neglect ionization which can prevent from reaching the ideal gas limit in real fluids~\cite{WhitePRR2024}) and thus $x\simeq 0$ and no solid component is present; (ii) at sufficiently low temperatures, where the system is in the solid state $x=1$ and there is no gas component. The melting temperature $T_{\rm m}$ is an important reference point as we should have $x=1$ at $T<T_{\rm m}$ and $x<1$ at $T>T_{\rm m}$. It is convenient to consider the rigidity parameter as a function of the relative temperature $T/T_{\rm m}$. At temperatures above the melting temperature, there are no temperature scales involved. The liquid rigidity parameter is expected to monotonically decay to zero~\cite{LinJCP2003,MoonPRR2024} and demonstrate {\it scale-invariant behavior}. The most general function satisfying these properties is a power-law function of temperature for temperatures above the melting point. We arrive at the following model for the liquid rigidity parameter:
\begin{equation}\label{rigidity}
x=
\begin{cases}
  1 & \mbox{ if $T<T_{\rm m}$}\\
  \left(T_{\rm m}/T\right)^{\alpha} & \mbox{ if $T\geq T_{\rm m}$}.
  \end{cases}
\end{equation}

The proposed behavior of the rigidity parameter is fully consistent with the analysis based on instantaneous normal modes and the instability factor IF reported in Ref.~\cite{MoonPRR2024}. In this study, molecular dynamics simulations were performed in the canonical NVT ensemble to follow the transition from solid to gas in argon, silicon, and iron. The initial state point was at 1 K and in the solid state (FCC lattice for argon and silicon and BCC lattice for iron). The lattice constants were 5.269 {\AA} in argon, 5.431 {\AA} in silicon, and 2.867 {\AA} in iron. Simulations were performed at temperatures of up to 10$^8$ K in argon, 10$^7$ K in silicon, and 10$^6$ K in iron. These extremely high temperatures were necessary to reach the ideal gas limit. The interatomic interactions corresponded to the Lennard-Jones potential for argon, Stillinger-Weber potential for silicon, and modified Johnson potential for iron. For further details on these simulations, see~\cite{MoonPRR2024}. The instantaneous normal mode densities obtained in the simulations were used to calculate the IF factors. Figure~\ref{Fig1} shows the comparison between the numerical results and our Eq.~(\ref{rigidity}). It represents a convincing proof of our concept.  

Substituting Eq.~(\ref{rigidity}) into Eq.~(\ref{c_v}) we immediately get
\begin{equation}\label{RT}
   c_{\rm v}=\frac{3}{2}+\frac{3}{2}(1-\alpha)\left(\frac{T_{\rm m}}{T}\right)^{\alpha}. 
\end{equation}
This coincides with the generalized RT scaling of the heat capacity in dense simple liquids proposed in Ref.~\cite{KhrapakPOF11_2024}, except {\it freezing} temperature was used for normalization there. Since we focus on the pure liquid phase properties here, the latter choice seems more appropriate. The liquid-solid coexistence is beyond the scope of this paper. Hence, the melting temperature refers to the temperature at the fluid boundary of fluid-solid co-existence. The exponent $\alpha$ is often close to $1/3$, but can be weakly material- and density-dependent~\cite{KhrapakPOF11_2024}. As a remark, we note that if Eq.~(\ref{RT}) with $\alpha=1/3$ is accepted, then condition $c_{\rm v}=2$ corresponding to the onset of rigid liquid (also known as the Frenkel line in the phase diagram)~\cite{TrachenkoBook,BrazhkinPRE2012,BrazhkinPRL2013} is met at $T\sim 10 T_{\rm m}$. According to Fig.~\ref{Fig1}, this roughly corresponds to $x\simeq 1-x\simeq 0.5$ as expected. 

\begin{figure}
\includegraphics[width=8cm]{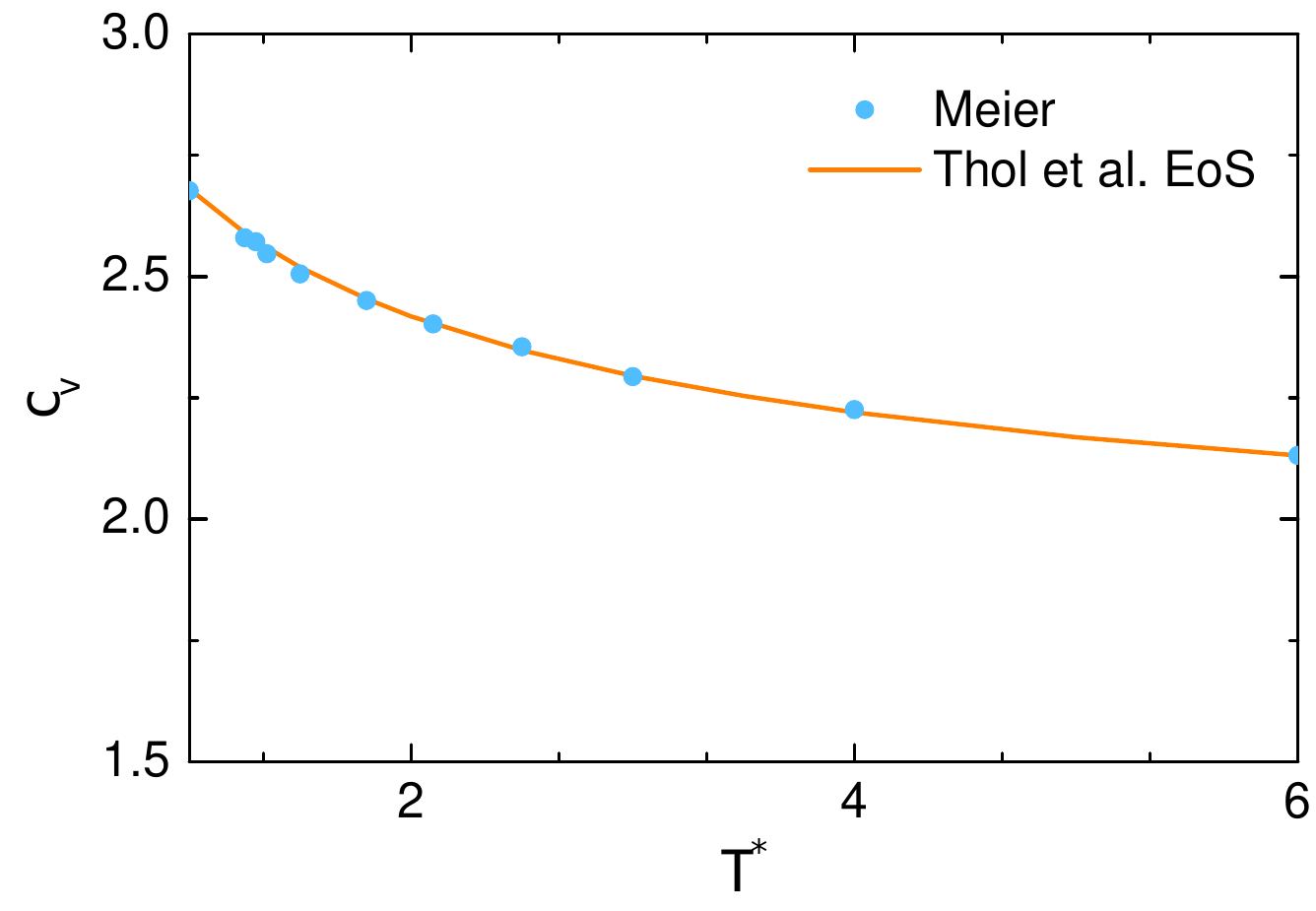}
\caption{(Color online) Reduced heat capacity of the LJ fluid along an isochore $\rho^*=0.9$. The solid line is calculated from the EoS in Ref.~\cite{Thol2016}.  Symbols correspond to the results tabulated in Ref.~\cite{Meier2002}. }
\label{Fig2}
\end{figure}

The applicability of generalized RT scaling to several model and real fluids has been discussed in Ref.~\cite{KhrapakPOF11_2024} and we do not repeat this here. For illustration purposes, we consider its performance in the conventional Lennard-Jones (LJ) liquid. Because freezing-temperature scaling is involved, we consider densities above the gas-liquid-solid triple point. The latter is located at $\rho^*\simeq 0.846$ and $T^*\simeq 0.694$, where $\rho^*=\rho\sigma^3$ and $T^*=T/\epsilon$ are conventional reduced LJ units of density and temperature~\cite{SousaJCP2012}. This explains what dense liquid means in our case.   

We calculate the heat capacity in the LJ fluid using the equation of state by Thol {\it et al.}.~\cite{Thol2016}. First, to verify its consistency, we compare its performance with the results tabulated by Meier~\cite{Meier2002} for
an isochore $\rho^*=0.9$. This comparison is shown in Fig.~\ref{Fig2}. The agreement is excellent. 

\begin{figure}
\includegraphics[width=8cm]{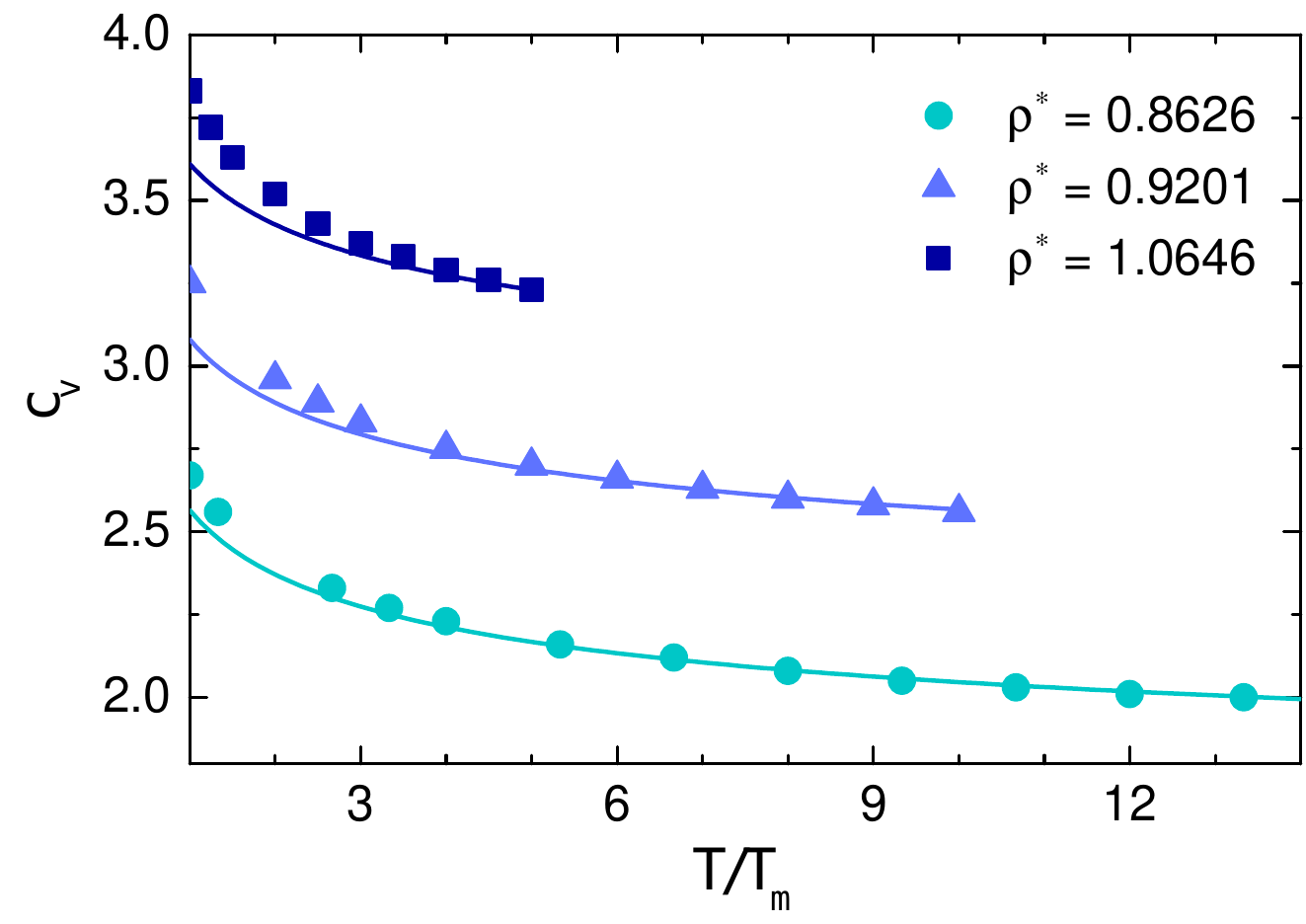}
\caption{(Color online) Reduced heat capacity of the LJ fluid along three isochores, $\rho^*=0.8626, 0.9201$, and $1.0646$, corresponding to the freezing densities at temperatures $T^*=0.75, 1$ and $2$, respectively (from bottom to top). Symbols refer to the calculation using an EoS from Ref.~\cite{Thol2016}. Curves are plotted using Eq.~(\ref{RT}) with $\alpha=0.29$, $0.28$, and $0.26$ (from bottom to top). The symbols and curves for different isochores are shifted by $0.5$ vertically for clarity. }
\label{Fig3}
\end{figure}

The calculated heat capacities of the LJ liquid along three isochores, $\rho^*=0.8626, 0.9201$, and $1.0646$, which correspond to the freezing densities at temperatures $T^*=0.75, 1$ and $2$, respectively~\cite{SousaJCP2012} are shown by symbols in Fig.~\ref{Fig3}. The curves correspond to Eq.~(\ref{RT}) with exponents $\alpha=0.29$, $0.28$, and $0.26$ (from bottom to top). The agreement is reasonably good, except for the close vicinity of the fluid-solid phase transition, where deviations can be observed. The deviations are probably due to the neglect of anharmonic effects, but this requires further attention. The agreement in the high-temperature limit is not surprising since this (ideal gas) limit is trivial. The simplicity of the model probably justifies some inaccuracy at lower temperatures. Note that the data sets for different isochores are shifted upwards by $0.5$ for clarity. 

In summary, we have shown that the generalized RT scaling $C_V^{\rm ex}\propto T^{-\alpha}$ can be easily derived from the two-phase model of liquid thermal energy, taking into account that the abundances of the gas and solid constituents should be described by a scale-free function of the reduced temperature. The exponent $\alpha$ is often close to $1/3$, but the exact value cannot be determined within the model and should be taken from empirical evidence.

\bibliography{SE_Ref}

\providecommand{\noopsort}[1]{}\providecommand{\singleletter}[1]{#1}%
\begin{thebibliography}{34}%
\makeatletter
\providecommand \@ifxundefined [1]{%
 \@ifx{#1\undefined}
}%
\providecommand \@ifnum [1]{%
 \ifnum #1\expandafter \@firstoftwo
 \else \expandafter \@secondoftwo
 \fi
}%
\providecommand \@ifx [1]{%
 \ifx #1\expandafter \@firstoftwo
 \else \expandafter \@secondoftwo
 \fi
}%
\providecommand \natexlab [1]{#1}%
\providecommand \enquote  [1]{``#1''}%
\providecommand \bibnamefont  [1]{#1}%
\providecommand \bibfnamefont [1]{#1}%
\providecommand \citenamefont [1]{#1}%
\providecommand \href@noop [0]{\@secondoftwo}%
\providecommand \href [0]{\begingroup \@sanitize@url \@href}%
\providecommand \@href[1]{\@@startlink{#1}\@@href}%
\providecommand \@@href[1]{\endgroup#1\@@endlink}%
\providecommand \@sanitize@url [0]{\catcode `\\12\catcode `\$12\catcode
  `\&12\catcode `\#12\catcode `\^12\catcode `\_12\catcode `\%12\relax}%
\providecommand \@@startlink[1]{}%
\providecommand \@@endlink[0]{}%
\providecommand \url  [0]{\begingroup\@sanitize@url \@url }%
\providecommand \@url [1]{\endgroup\@href {#1}{\urlprefix }}%
\providecommand \urlprefix  [0]{URL }%
\providecommand \Eprint [0]{\href }%
\providecommand \doibase [0]{http://dx.doi.org/}%
\providecommand \selectlanguage [0]{\@gobble}%
\providecommand \bibinfo  [0]{\@secondoftwo}%
\providecommand \bibfield  [0]{\@secondoftwo}%
\providecommand \translation [1]{[#1]}%
\providecommand \BibitemOpen [0]{}%
\providecommand \bibitemStop [0]{}%
\providecommand \bibitemNoStop [0]{.\EOS\space}%
\providecommand \EOS [0]{\spacefactor3000\relax}%
\providecommand \BibitemShut  [1]{\csname bibitem#1\endcsname}%
\let\auto@bib@innerbib\@empty
%</preamble>
\bibitem [{\citenamefont {Trachenko}(2023)}]{TrachenkoBook}%
  \BibitemOpen
  \bibfield  {author} {\bibinfo {author} {\bibfnamefont {K.}~\bibnamefont
  {Trachenko}},\ }\href@noop {} {\emph {\bibinfo {title} {Theory of liquids:
  {F}rom excitations to Thermodynamics}}}\ (\bibinfo  {publisher} {Cambridge
  University Press},\ \bibinfo {address} {Cambridge, England},\ \bibinfo {year}
  {2023})\BibitemShut {NoStop}%
\bibitem [{\citenamefont {Landau}\ \emph {et~al.}(1980)\citenamefont {Landau},
  \citenamefont {Lifshic},\ and\ \citenamefont {Pitaevskii}}]{LandauStatPhys}%
  \BibitemOpen
  \bibfield  {author} {\bibinfo {author} {\bibfnamefont {L.~D.}\ \bibnamefont
  {Landau}}, \bibinfo {author} {\bibfnamefont {E.~M.}\ \bibnamefont {Lifshic}},
  \ and\ \bibinfo {author} {\bibfnamefont {L.~P.}\ \bibnamefont {Pitaevskii}},\
  }\href@noop {} {\emph {\bibinfo {title} {Statistical Physics}}}\ (\bibinfo
  {publisher} {Butterworth-Heinemann},\ \bibinfo {address} {Oxford},\ \bibinfo
  {year} {1980})\BibitemShut {NoStop}%
\bibitem [{\citenamefont {Rosenfeld}\ and\ \citenamefont
  {Tarazona}(1998)}]{RosenfeldMolPhys1998}%
  \BibitemOpen
  \bibfield  {author} {\bibinfo {author} {\bibfnamefont {Y.}~\bibnamefont
  {Rosenfeld}}\ and\ \bibinfo {author} {\bibfnamefont {P.}~\bibnamefont
  {Tarazona}},\ }\bibfield  {title} {\enquote {\bibinfo {title} {Density
  functional theory and the asymptotic high density expansion of the free
  energy of classical solids and fluids},}\ }\href {\doibase
  10.1080/00268979809483145} {\bibfield  {journal} {\bibinfo  {journal} {Mol.
  Phys.}\ }\textbf {\bibinfo {volume} {95}},\ \bibinfo {pages} {141--150}
  (\bibinfo {year} {1998})}\BibitemShut {NoStop}%
\bibitem [{\citenamefont {Rosenfeld}(2000)}]{RosenfeldPRE2000}%
  \BibitemOpen
  \bibfield  {author} {\bibinfo {author} {\bibfnamefont {Y.}~\bibnamefont
  {Rosenfeld}},\ }\bibfield  {title} {\enquote {\bibinfo {title}
  {Excess-entropy and freezing-temperature scalings for transport coefficients:
  Self-diffusion in {Y}ukawa systems},}\ }\href {\doibase
  10.1103/physreve.62.7524} {\bibfield  {journal} {\bibinfo  {journal} {Phys.
  Rev. E}\ }\textbf {\bibinfo {volume} {62}},\ \bibinfo {pages} {7524--7527}
  (\bibinfo {year} {2000})}\BibitemShut {NoStop}%
\bibitem [{\citenamefont {Ingebrigtsen}\ \emph {et~al.}(2013)\citenamefont
  {Ingebrigtsen}, \citenamefont {Veldhorst}, \citenamefont {Schr{\o}der},\ and\
  \citenamefont {Dyre}}]{IngebrigtsenJCP2013}%
  \BibitemOpen
  \bibfield  {author} {\bibinfo {author} {\bibfnamefont {T.~S.}\ \bibnamefont
  {Ingebrigtsen}}, \bibinfo {author} {\bibfnamefont {A.~A.}\ \bibnamefont
  {Veldhorst}}, \bibinfo {author} {\bibfnamefont {T.~B.}\ \bibnamefont
  {Schr{\o}der}}, \ and\ \bibinfo {author} {\bibfnamefont {J.~C.}\ \bibnamefont
  {Dyre}},\ }\bibfield  {title} {\enquote {\bibinfo {title} {Communication: The
  {R}osenfeld-{T}arazona expression for liquids' specific heat: A numerical
  investigation of eighteen systems},}\ }\href {\doibase 10.1063/1.4827865}
  {\bibfield  {journal} {\bibinfo  {journal} {J. Chem. Phys.}\ }\textbf
  {\bibinfo {volume} {139}},\ \bibinfo {pages} {171101} (\bibinfo {year}
  {2013})}\BibitemShut {NoStop}%
\bibitem [{\citenamefont {Ingebrigtsen}\ \emph {et~al.}(2012)\citenamefont
  {Ingebrigtsen}, \citenamefont {Schrøder},\ and\ \citenamefont
  {Dyre}}]{IngebrigtsenPRX2012}%
  \BibitemOpen
  \bibfield  {author} {\bibinfo {author} {\bibfnamefont {T.~S.}\ \bibnamefont
  {Ingebrigtsen}}, \bibinfo {author} {\bibfnamefont {T.~B.}\ \bibnamefont
  {Schrøder}}, \ and\ \bibinfo {author} {\bibfnamefont {J.~C.}\ \bibnamefont
  {Dyre}},\ }\bibfield  {title} {\enquote {\bibinfo {title} {What is a simple
  liquid?}}\ }\href {\doibase 10.1103/physrevx.2.011011} {\bibfield  {journal}
  {\bibinfo  {journal} {Phys. Rev. X}\ }\textbf {\bibinfo {volume} {2}},\
  \bibinfo {pages} {011011} (\bibinfo {year} {2012})}\BibitemShut {NoStop}%
\bibitem [{\citenamefont {Dyre}(2014)}]{DyreJPCB2014}%
  \BibitemOpen
  \bibfield  {author} {\bibinfo {author} {\bibfnamefont {J.~C.}\ \bibnamefont
  {Dyre}},\ }\bibfield  {title} {\enquote {\bibinfo {title} {Hidden scale
  invariance in condensed matter},}\ }\href {\doibase 10.1021/jp501852b}
  {\bibfield  {journal} {\bibinfo  {journal} {J. Phys. Chem. B}\ }\textbf
  {\bibinfo {volume} {118}},\ \bibinfo {pages} {10007--10024} (\bibinfo {year}
  {2014})}\BibitemShut {NoStop}%
\bibitem [{\citenamefont {Khrapak}(2024{\natexlab{a}})}]{KhrapakPRE09_2024}%
  \BibitemOpen
  \bibfield  {author} {\bibinfo {author} {\bibfnamefont {S.~A.}\ \bibnamefont
  {Khrapak}},\ }\bibfield  {title} {\enquote {\bibinfo {title} {Entropy of
  strongly coupled {Y}ukawa fluids},}\ }\href {\doibase
  10.1103/physreve.110.034602} {\bibfield  {journal} {\bibinfo  {journal}
  {Phys. Rev. E}\ }\textbf {\bibinfo {volume} {110}},\ \bibinfo {pages}
  {034602} (\bibinfo {year} {2024}{\natexlab{a}})}\BibitemShut {NoStop}%
\bibitem [{\citenamefont {Khrapak}\ and\ \citenamefont
  {Khrapak}(2024)}]{KhrapakPOF11_2024}%
  \BibitemOpen
  \bibfield  {author} {\bibinfo {author} {\bibfnamefont {S.~A.}\ \bibnamefont
  {Khrapak}}\ and\ \bibinfo {author} {\bibfnamefont {A.~G.}\ \bibnamefont
  {Khrapak}},\ }\bibfield  {title} {\enquote {\bibinfo {title} {Generalized
  {R}osenfeld--{T}arazona scaling and high-density specific heat of simple
  liquids},}\ }\href {\doibase 10.1063/5.0230219} {\bibfield  {journal}
  {\bibinfo  {journal} {Phys. Fluids}\ }\textbf {\bibinfo {volume} {36}},\
  \bibinfo {pages} {117119} (\bibinfo {year} {2024})}\BibitemShut {NoStop}%
\bibitem [{\citenamefont {Trachenko}(2008)}]{TrachenkoPRB2008}%
  \BibitemOpen
  \bibfield  {author} {\bibinfo {author} {\bibfnamefont {K.}~\bibnamefont
  {Trachenko}},\ }\bibfield  {title} {\enquote {\bibinfo {title} {Heat capacity
  of liquids: {A}n approach from the solid phase},}\ }\href {\doibase
  10.1103/physrevb.78.104201} {\bibfield  {journal} {\bibinfo  {journal} {Phys.
  Rev. B}\ }\textbf {\bibinfo {volume} {78}},\ \bibinfo {pages} {104201}
  (\bibinfo {year} {2008})}\BibitemShut {NoStop}%
\bibitem [{\citenamefont {Bolmatov}\ \emph {et~al.}(2012)\citenamefont
  {Bolmatov}, \citenamefont {Brazhkin},\ and\ \citenamefont
  {Trachenko}}]{BolmatovSciRep2012}%
  \BibitemOpen
  \bibfield  {author} {\bibinfo {author} {\bibfnamefont {D.}~\bibnamefont
  {Bolmatov}}, \bibinfo {author} {\bibfnamefont {V.~V.}\ \bibnamefont
  {Brazhkin}}, \ and\ \bibinfo {author} {\bibfnamefont {K.}~\bibnamefont
  {Trachenko}},\ }\bibfield  {title} {\enquote {\bibinfo {title} {The phonon
  theory of liquid thermodynamics},}\ }\href {\doibase 10.1038/srep00421}
  {\bibfield  {journal} {\bibinfo  {journal} {Sci. Rep.}\ }\textbf {\bibinfo
  {volume} {2}},\ \bibinfo {pages} {421} (\bibinfo {year} {2012})}\BibitemShut
  {NoStop}%
\bibitem [{\citenamefont {Bolmatov}\ \emph {et~al.}(2015)\citenamefont
  {Bolmatov}, \citenamefont {Zav'yalov}, \citenamefont {Zhernenkov},
  \citenamefont {Musaev},\ and\ \citenamefont {Cai}}]{BolmatovAP2015}%
  \BibitemOpen
  \bibfield  {author} {\bibinfo {author} {\bibfnamefont {D.}~\bibnamefont
  {Bolmatov}}, \bibinfo {author} {\bibfnamefont {D.}~\bibnamefont {Zav'yalov}},
  \bibinfo {author} {\bibfnamefont {M.}~\bibnamefont {Zhernenkov}}, \bibinfo
  {author} {\bibfnamefont {E.~T.}\ \bibnamefont {Musaev}}, \ and\ \bibinfo
  {author} {\bibfnamefont {Y.~Q.}\ \bibnamefont {Cai}},\ }\bibfield  {title}
  {\enquote {\bibinfo {title} {Unified phonon-based approach to the
  thermodynamics of solid, liquid and gas states},}\ }\href {\doibase
  10.1016/j.aop.2015.09.018} {\bibfield  {journal} {\bibinfo  {journal} {Annals
  of Phys.}\ }\textbf {\bibinfo {volume} {363}},\ \bibinfo {pages} {221--242}
  (\bibinfo {year} {2015})}\BibitemShut {NoStop}%
\bibitem [{\citenamefont {Trachenko}\ and\ \citenamefont
  {Brazhkin}(2015)}]{TrachenkoRPP2015}%
  \BibitemOpen
  \bibfield  {author} {\bibinfo {author} {\bibfnamefont {K.}~\bibnamefont
  {Trachenko}}\ and\ \bibinfo {author} {\bibfnamefont {V.~V.}\ \bibnamefont
  {Brazhkin}},\ }\bibfield  {title} {\enquote {\bibinfo {title} {Collective
  modes and thermodynamics of the liquid state},}\ }\href {\doibase
  10.1088/0034-4885/79/1/016502} {\bibfield  {journal} {\bibinfo  {journal}
  {Rep. Progr. Phys.}\ }\textbf {\bibinfo {volume} {79}},\ \bibinfo {pages}
  {016502} (\bibinfo {year} {2015})}\BibitemShut {NoStop}%
\bibitem [{\citenamefont {Bolmatov}(2022)}]{BolmatovJPCL2022}%
  \BibitemOpen
  \bibfield  {author} {\bibinfo {author} {\bibfnamefont {D.}~\bibnamefont
  {Bolmatov}},\ }\bibfield  {title} {\enquote {\bibinfo {title} {The phonon
  theory of liquids and biological fluids: {D}evelopments and applications},}\
  }\href {\doibase 10.1021/acs.jpclett.2c01779} {\bibfield  {journal} {\bibinfo
   {journal} {J. Phys. Chem. Lett.}\ }\textbf {\bibinfo {volume} {13}},\
  \bibinfo {pages} {7121–7129} (\bibinfo {year} {2022})}\BibitemShut
  {NoStop}%
\bibitem [{\citenamefont {Liu}\ and\ \citenamefont
  {Baggioli}(2025)}]{LiuPRB2025}%
  \BibitemOpen
  \bibfield  {author} {\bibinfo {author} {\bibfnamefont {Y.}~\bibnamefont
  {Liu}}\ and\ \bibinfo {author} {\bibfnamefont {M.}~\bibnamefont {Baggioli}},\
  }\bibfield  {title} {\enquote {\bibinfo {title} {Revisiting the phonon theory
  of liquid heat capacity: {L}ow-frequency shear modes and intramolecular
  vibrations},}\ }\href {\doibase 10.1103/physrevb.111.144201} {\bibfield
  {journal} {\bibinfo  {journal} {Phys. Rev. B}\ }\textbf {\bibinfo {volume}
  {111}},\ \bibinfo {pages} {144201} (\bibinfo {year} {2025})}\BibitemShut
  {NoStop}%
\bibitem [{\citenamefont {Frenkel}(1955)}]{FrenkelBook}%
  \BibitemOpen
  \bibfield  {author} {\bibinfo {author} {\bibfnamefont {Y.}~\bibnamefont
  {Frenkel}},\ }\href {https://cds.cern.ch/record/106808} {\emph {\bibinfo
  {title} {{Kinetic theory of liquids}}}}\ (\bibinfo  {publisher} {Dover},\
  \bibinfo {address} {New York, NY},\ \bibinfo {year} {1955})\BibitemShut
  {NoStop}%
\bibitem [{\citenamefont {Khrapak}(2024{\natexlab{b}})}]{KhrapakPhysRep2024}%
  \BibitemOpen
  \bibfield  {author} {\bibinfo {author} {\bibfnamefont {S.A.}\ \bibnamefont
  {Khrapak}},\ }\bibfield  {title} {\enquote {\bibinfo {title} {Elementary
  vibrational model for transport properties of dense fluids},}\ }\href
  {\doibase 10.1016/j.physrep.2023.11.004} {\bibfield  {journal} {\bibinfo
  {journal} {Phys. Rep.}\ }\textbf {\bibinfo {volume} {1050}},\ \bibinfo
  {pages} {1} (\bibinfo {year} {2024}{\natexlab{b}})}\BibitemShut {NoStop}%
\bibitem [{\citenamefont {Khrapak}\ \emph {et~al.}(2019)\citenamefont
  {Khrapak}, \citenamefont {Khrapak}, \citenamefont {Kryuchkov},\ and\
  \citenamefont {Yurchenko}}]{KhrapakJCP2019}%
  \BibitemOpen
  \bibfield  {author} {\bibinfo {author} {\bibfnamefont {S.~A.}\ \bibnamefont
  {Khrapak}}, \bibinfo {author} {\bibfnamefont {A.~G.}\ \bibnamefont
  {Khrapak}}, \bibinfo {author} {\bibfnamefont {N.~P.}\ \bibnamefont
  {Kryuchkov}}, \ and\ \bibinfo {author} {\bibfnamefont {S.~O.}\ \bibnamefont
  {Yurchenko}},\ }\bibfield  {title} {\enquote {\bibinfo {title} {Onset of
  transverse (shear) waves in strongly-coupled {Y}ukawa fluids},}\ }\href
  {\doibase 10.1063/1.5088141} {\bibfield  {journal} {\bibinfo  {journal} {J.
  Chem. Phys.}\ }\textbf {\bibinfo {volume} {150}},\ \bibinfo {pages} {104503}
  (\bibinfo {year} {2019})}\BibitemShut {NoStop}%
\bibitem [{\citenamefont {Kryuchkov}\ \emph
  {et~al.}(2019{\natexlab{a}})\citenamefont {Kryuchkov}, \citenamefont
  {Mistryukova}, \citenamefont {Brazhkin},\ and\ \citenamefont
  {Yurchenko}}]{KryuchkovSciRep2019}%
  \BibitemOpen
  \bibfield  {author} {\bibinfo {author} {\bibfnamefont {N.~P.}\ \bibnamefont
  {Kryuchkov}}, \bibinfo {author} {\bibfnamefont {L.~A.}\ \bibnamefont
  {Mistryukova}}, \bibinfo {author} {\bibfnamefont {V.~V.}\ \bibnamefont
  {Brazhkin}}, \ and\ \bibinfo {author} {\bibfnamefont {S.~O.}\ \bibnamefont
  {Yurchenko}},\ }\bibfield  {title} {\enquote {\bibinfo {title} {Excitation
  spectra in fluids: How to analyze them properly},}\ }\href {\doibase
  10.1038/s41598-019-46979-y} {\bibfield  {journal} {\bibinfo  {journal} {Sci.
  Rep.}\ }\textbf {\bibinfo {volume} {9}},\ \bibinfo {pages} {10483} (\bibinfo
  {year} {2019}{\natexlab{a}})}\BibitemShut {NoStop}%
\bibitem [{\citenamefont {Kryuchkov}\ \emph
  {et~al.}(2019{\natexlab{b}})\citenamefont {Kryuchkov}, \citenamefont
  {Brazhkin},\ and\ \citenamefont {Yurchenko}}]{KryuchkovJCPL2019}%
  \BibitemOpen
  \bibfield  {author} {\bibinfo {author} {\bibfnamefont {N.~P.}\ \bibnamefont
  {Kryuchkov}}, \bibinfo {author} {\bibfnamefont {V.~V.}\ \bibnamefont
  {Brazhkin}}, \ and\ \bibinfo {author} {\bibfnamefont {S.~O.}\ \bibnamefont
  {Yurchenko}},\ }\bibfield  {title} {\enquote {\bibinfo {title} {Anticrossing
  of longitudinal and transverse modes in simple fluids},}\ }\href {\doibase
  10.1021/acs.jpclett.9b01468} {\bibfield  {journal} {\bibinfo  {journal} {J.
  Phys. Chem. Lett.}\ }\textbf {\bibinfo {volume} {10}},\ \bibinfo {pages}
  {4470–4475} (\bibinfo {year} {2019}{\natexlab{b}})}\BibitemShut {NoStop}%
\bibitem [{\citenamefont {Yakovlev}\ \emph {et~al.}(2020)\citenamefont
  {Yakovlev}, \citenamefont {Kryuchkov}, \citenamefont {Ovcharov},
  \citenamefont {Sapelkin}, \citenamefont {Brazhkin},\ and\ \citenamefont
  {Yurchenko}}]{YakovlevJCPL2020}%
  \BibitemOpen
  \bibfield  {author} {\bibinfo {author} {\bibfnamefont {E.~V.}\ \bibnamefont
  {Yakovlev}}, \bibinfo {author} {\bibfnamefont {N.~P.}\ \bibnamefont
  {Kryuchkov}}, \bibinfo {author} {\bibfnamefont {P.~V.}\ \bibnamefont
  {Ovcharov}}, \bibinfo {author} {\bibfnamefont {A.~V.}\ \bibnamefont
  {Sapelkin}}, \bibinfo {author} {\bibfnamefont {V.~V.}\ \bibnamefont
  {Brazhkin}}, \ and\ \bibinfo {author} {\bibfnamefont {S.~O.}\ \bibnamefont
  {Yurchenko}},\ }\bibfield  {title} {\enquote {\bibinfo {title} {Direct
  experimental evidence of longitudinal and transverse mode hybridization and
  anticrossing in simple model fluids},}\ }\href {\doibase
  10.1021/acs.jpclett.9b03568} {\bibfield  {journal} {\bibinfo  {journal} {J.
  Phys. Chem. Lett.}\ }\textbf {\bibinfo {volume} {11}},\ \bibinfo {pages}
  {1370–1376} (\bibinfo {year} {2020})}\BibitemShut {NoStop}%
\bibitem [{\citenamefont {Khrapak}\ and\ \citenamefont
  {Couedel}(2020)}]{KhrapakPRE09_2020}%
  \BibitemOpen
  \bibfield  {author} {\bibinfo {author} {\bibfnamefont {S.}~\bibnamefont
  {Khrapak}}\ and\ \bibinfo {author} {\bibfnamefont {L.}~\bibnamefont
  {Couedel}},\ }\bibfield  {title} {\enquote {\bibinfo {title} {Dispersion
  relations of {Y}ukawa fluids at weak and moderate coupling},}\ }\href
  {\doibase 10.1103/physreve.102.033207} {\bibfield  {journal} {\bibinfo
  {journal} {Phys. Rev. E}\ }\textbf {\bibinfo {volume} {102}},\ \bibinfo
  {pages} {033207} (\bibinfo {year} {2020})}\BibitemShut {NoStop}%
\bibitem [{\citenamefont {Kryuchkov}\ \emph {et~al.}(2020)\citenamefont
  {Kryuchkov}, \citenamefont {Mistryukova}, \citenamefont {Sapelkin},
  \citenamefont {Brazhkin},\ and\ \citenamefont
  {Yurchenko}}]{KryuchkovPRL2020}%
  \BibitemOpen
  \bibfield  {author} {\bibinfo {author} {\bibfnamefont {N.~P.}\ \bibnamefont
  {Kryuchkov}}, \bibinfo {author} {\bibfnamefont {L.~A.}\ \bibnamefont
  {Mistryukova}}, \bibinfo {author} {\bibfnamefont {A.~V.}\ \bibnamefont
  {Sapelkin}}, \bibinfo {author} {\bibfnamefont {V.~V.}\ \bibnamefont
  {Brazhkin}}, \ and\ \bibinfo {author} {\bibfnamefont {S.~O.}\ \bibnamefont
  {Yurchenko}},\ }\bibfield  {title} {\enquote {\bibinfo {title} {Universal
  effect of excitation dispersion on the heat capacity and gapped states in
  fluids},}\ }\href {\doibase 10.1103/physrevlett.125.125501} {\bibfield
  {journal} {\bibinfo  {journal} {Phys. Rev. Lett.}\ }\textbf {\bibinfo
  {volume} {125}},\ \bibinfo {pages} {125501} (\bibinfo {year}
  {2020})}\BibitemShut {NoStop}%
\bibitem [{\citenamefont {Bryk}\ and\ \citenamefont
  {Ruocco}(2025)}]{BrykPRE2025}%
  \BibitemOpen
  \bibfield  {author} {\bibinfo {author} {\bibfnamefont {T.}~\bibnamefont
  {Bryk}}\ and\ \bibinfo {author} {\bibfnamefont {G.}~\bibnamefont {Ruocco}},\
  }\bibfield  {title} {\enquote {\bibinfo {title} {Transverse collective
  excitations in liquids: {I}s the phonon theory of liquid thermodynamics
  correct?}}\ }\href {\doibase 10.1103/physreve.111.064102} {\bibfield
  {journal} {\bibinfo  {journal} {Phys. Rev. E}\ }\textbf {\bibinfo {volume}
  {111}},\ \bibinfo {pages} {064102} (\bibinfo {year} {2025})}\BibitemShut
  {NoStop}%
\bibitem [{\citenamefont {Lin}\ \emph {et~al.}(2003)\citenamefont {Lin},
  \citenamefont {Blanco},\ and\ \citenamefont {Goddard}}]{LinJCP2003}%
  \BibitemOpen
  \bibfield  {author} {\bibinfo {author} {\bibfnamefont {S.-T.}\ \bibnamefont
  {Lin}}, \bibinfo {author} {\bibfnamefont {M.}~\bibnamefont {Blanco}}, \ and\
  \bibinfo {author} {\bibfnamefont {W.~A.}\ \bibnamefont {Goddard}},\
  }\bibfield  {title} {\enquote {\bibinfo {title} {The two-phase model for
  calculating thermodynamic properties of liquids from molecular dynamics:
  {V}alidation for the phase diagram of {L}ennard-{J}ones fluids},}\ }\href
  {\doibase 10.1063/1.1624057} {\bibfield  {journal} {\bibinfo  {journal} {J.
  Chem. Phys.}\ }\textbf {\bibinfo {volume} {119}},\ \bibinfo {pages}
  {11792–11805} (\bibinfo {year} {2003})}\BibitemShut {NoStop}%
\bibitem [{\citenamefont {Pascal}\ \emph {et~al.}(2011)\citenamefont {Pascal},
  \citenamefont {Lin},\ and\ \citenamefont {Goddard~III}}]{PascalPCCP2011}%
  \BibitemOpen
  \bibfield  {author} {\bibinfo {author} {\bibfnamefont {T.~A.}\ \bibnamefont
  {Pascal}}, \bibinfo {author} {\bibfnamefont {S.-T.}\ \bibnamefont {Lin}}, \
  and\ \bibinfo {author} {\bibfnamefont {W.~A.}\ \bibnamefont {Goddard~III}},\
  }\bibfield  {title} {\enquote {\bibinfo {title} {Thermodynamics of liquids:
  {S}tandard molar entropies and heat capacities of common solvents from {2PT}
  molecular dynamics},}\ }\href {\doibase 10.1039/c0cp01549k} {\bibfield
  {journal} {\bibinfo  {journal} {Phys. Chem. Chem. Phys.}\ }\textbf {\bibinfo
  {volume} {13}},\ \bibinfo {pages} {169–181} (\bibinfo {year}
  {2011})}\BibitemShut {NoStop}%
\bibitem [{\citenamefont {Moon}\ \emph {et~al.}(2024)\citenamefont {Moon},
  \citenamefont {Thébaud}, \citenamefont {Lindsay},\ and\ \citenamefont
  {Egami}}]{MoonPRR2024}%
  \BibitemOpen
  \bibfield  {author} {\bibinfo {author} {\bibfnamefont {J.}~\bibnamefont
  {Moon}}, \bibinfo {author} {\bibfnamefont {S.}~\bibnamefont {Thébaud}},
  \bibinfo {author} {\bibfnamefont {L.}~\bibnamefont {Lindsay}}, \ and\
  \bibinfo {author} {\bibfnamefont {T.}~\bibnamefont {Egami}},\ }\bibfield
  {title} {\enquote {\bibinfo {title} {Normal mode description of phases of
  matter: {A}pplication to heat capacity},}\ }\href {\doibase
  10.1103/physrevresearch.6.013206} {\bibfield  {journal} {\bibinfo  {journal}
  {Phys. Rev. Research}\ }\textbf {\bibinfo {volume} {6}},\ \bibinfo {pages}
  {013206} (\bibinfo {year} {2024})}\BibitemShut {NoStop}%
\bibitem [{\citenamefont {Khrapak}(2025)}]{KhrapakPRE2025}%
  \BibitemOpen
  \bibfield  {author} {\bibinfo {author} {\bibfnamefont {S.~A.}\ \bibnamefont
  {Khrapak}},\ }\bibfield  {title} {\enquote {\bibinfo {title} {Speed of sound
  in dense simple liquids},}\ }\href {\doibase 10.1103/5dtk-4x7m} {\bibfield
  {journal} {\bibinfo  {journal} {Phys. Rev. E}\ }\textbf {\bibinfo {volume}
  {111}},\ \bibinfo {pages} {065423} (\bibinfo {year} {2025})}\BibitemShut
  {NoStop}%
\bibitem [{\citenamefont {White}\ \emph {et~al.}(2024)\citenamefont {White},
  \citenamefont {Poole}, \citenamefont {McBride}, \citenamefont {Oliver},
  \citenamefont {Descamps}, \citenamefont {Fletcher}, \citenamefont
  {Angermeier}, \citenamefont {Allen}, \citenamefont {Appel}, \citenamefont
  {Condamine}, \citenamefont {Curry}, \citenamefont {Dallari}, \citenamefont
  {Funk}, \citenamefont {Galtier}, \citenamefont {Gamboa}, \citenamefont
  {Gauthier}, \citenamefont {Graham}, \citenamefont {Goede}, \citenamefont
  {Haden}, \citenamefont {Kim}, \citenamefont {Lee}, \citenamefont
  {Ofori-Okai}, \citenamefont {Richardson}, \citenamefont {Rigby},
  \citenamefont {Schoenwaelder}, \citenamefont {Sun}, \citenamefont {Witte},
  \citenamefont {Tschentscher}, \citenamefont {Zastrau}, \citenamefont
  {Nagler}, \citenamefont {Hastings}, \citenamefont {Monaco}, \citenamefont
  {Gericke}, \citenamefont {Glenzer},\ and\ \citenamefont
  {Gregori}}]{WhitePRR2024}%
  \BibitemOpen
  \bibfield  {author} {\bibinfo {author} {\bibfnamefont {T.}~\bibnamefont
  {White}}, \bibinfo {author} {\bibfnamefont {H.}~\bibnamefont {Poole}},
  \bibinfo {author} {\bibfnamefont {E.}~\bibnamefont {McBride}}, \bibinfo
  {author} {\bibfnamefont {M.}~\bibnamefont {Oliver}}, \bibinfo {author}
  {\bibfnamefont {A.}~\bibnamefont {Descamps}}, \bibinfo {author}
  {\bibfnamefont {L.}~\bibnamefont {Fletcher}}, \bibinfo {author}
  {\bibfnamefont {W.}~\bibnamefont {Angermeier}}, \bibinfo {author}
  {\bibfnamefont {C.}~\bibnamefont {Allen}}, \bibinfo {author} {\bibfnamefont
  {K.}~\bibnamefont {Appel}}, \bibinfo {author} {\bibfnamefont
  {F.}~\bibnamefont {Condamine}}, \bibinfo {author} {\bibfnamefont
  {C.}~\bibnamefont {Curry}}, \bibinfo {author} {\bibfnamefont
  {F.}~\bibnamefont {Dallari}}, \bibinfo {author} {\bibfnamefont
  {S.}~\bibnamefont {Funk}}, \bibinfo {author} {\bibfnamefont {E.}~\bibnamefont
  {Galtier}}, \bibinfo {author} {\bibfnamefont {E.}~\bibnamefont {Gamboa}},
  \bibinfo {author} {\bibfnamefont {M.}~\bibnamefont {Gauthier}}, \bibinfo
  {author} {\bibfnamefont {P.}~\bibnamefont {Graham}}, \bibinfo {author}
  {\bibfnamefont {S.}~\bibnamefont {Goede}}, \bibinfo {author} {\bibfnamefont
  {D.}~\bibnamefont {Haden}}, \bibinfo {author} {\bibfnamefont
  {J.}~\bibnamefont {Kim}}, \bibinfo {author} {\bibfnamefont {H.}~\bibnamefont
  {Lee}}, \bibinfo {author} {\bibfnamefont {B.}~\bibnamefont {Ofori-Okai}},
  \bibinfo {author} {\bibfnamefont {S.}~\bibnamefont {Richardson}}, \bibinfo
  {author} {\bibfnamefont {A.}~\bibnamefont {Rigby}}, \bibinfo {author}
  {\bibfnamefont {C.}~\bibnamefont {Schoenwaelder}}, \bibinfo {author}
  {\bibfnamefont {P.}~\bibnamefont {Sun}}, \bibinfo {author} {\bibfnamefont
  {B.}~\bibnamefont {Witte}}, \bibinfo {author} {\bibfnamefont
  {T.}~\bibnamefont {Tschentscher}}, \bibinfo {author} {\bibfnamefont
  {U.}~\bibnamefont {Zastrau}}, \bibinfo {author} {\bibfnamefont
  {B.}~\bibnamefont {Nagler}}, \bibinfo {author} {\bibfnamefont
  {J.}~\bibnamefont {Hastings}}, \bibinfo {author} {\bibfnamefont
  {G.}~\bibnamefont {Monaco}}, \bibinfo {author} {\bibfnamefont
  {D.}~\bibnamefont {Gericke}}, \bibinfo {author} {\bibfnamefont
  {S.}~\bibnamefont {Glenzer}}, \ and\ \bibinfo {author} {\bibfnamefont
  {G.}~\bibnamefont {Gregori}},\ }\bibfield  {title} {\enquote {\bibinfo
  {title} {Speed of sound in methane under conditions of planetary
  interiors},}\ }\href {\doibase 10.1103/physrevresearch.6.l022029} {\bibfield
  {journal} {\bibinfo  {journal} {Phys. Rev. Research}\ }\textbf {\bibinfo
  {volume} {6}},\ \bibinfo {pages} {L022029} (\bibinfo {year}
  {2024})}\BibitemShut {NoStop}%
\bibitem [{\citenamefont {Brazhkin}\ \emph {et~al.}(2012)\citenamefont
  {Brazhkin}, \citenamefont {Fomin}, \citenamefont {Lyapin}, \citenamefont
  {Ryzhov},\ and\ \citenamefont {Trachenko}}]{BrazhkinPRE2012}%
  \BibitemOpen
  \bibfield  {author} {\bibinfo {author} {\bibfnamefont {V.~V.}\ \bibnamefont
  {Brazhkin}}, \bibinfo {author} {\bibfnamefont {Yu.~D.}\ \bibnamefont
  {Fomin}}, \bibinfo {author} {\bibfnamefont {A.~G.}\ \bibnamefont {Lyapin}},
  \bibinfo {author} {\bibfnamefont {V.~N.}\ \bibnamefont {Ryzhov}}, \ and\
  \bibinfo {author} {\bibfnamefont {K.}~\bibnamefont {Trachenko}},\ }\bibfield
  {title} {\enquote {\bibinfo {title} {Two liquid states of matter: A dynamic
  line on a phase diagram},}\ }\href {\doibase 10.1103/physreve.85.031203}
  {\bibfield  {journal} {\bibinfo  {journal} {Phys. Rev. E}\ }\textbf {\bibinfo
  {volume} {85}},\ \bibinfo {pages} {031203} (\bibinfo {year}
  {2012})}\BibitemShut {NoStop}%
\bibitem [{\citenamefont {Brazhkin}\ \emph {et~al.}(2013)\citenamefont
  {Brazhkin}, \citenamefont {Fomin}, \citenamefont {Lyapin}, \citenamefont
  {Ryzhov}, \citenamefont {Tsiok},\ and\ \citenamefont
  {Trachenko}}]{BrazhkinPRL2013}%
  \BibitemOpen
  \bibfield  {author} {\bibinfo {author} {\bibfnamefont {V.~V.}\ \bibnamefont
  {Brazhkin}}, \bibinfo {author} {\bibfnamefont {Yu.~D.}\ \bibnamefont
  {Fomin}}, \bibinfo {author} {\bibfnamefont {A.~G.}\ \bibnamefont {Lyapin}},
  \bibinfo {author} {\bibfnamefont {V.~N.}\ \bibnamefont {Ryzhov}}, \bibinfo
  {author} {\bibfnamefont {E.~N.}\ \bibnamefont {Tsiok}}, \ and\ \bibinfo
  {author} {\bibfnamefont {K.}~\bibnamefont {Trachenko}},\ }\bibfield  {title}
  {\enquote {\bibinfo {title} {Liquid-gas'' transition in the supercritical
  region: {F}undamental changes in the particle dynamics},}\ }\href@noop {}
  {\bibfield  {journal} {\bibinfo  {journal} {Phys. Rev. Lett.}\ }\textbf
  {\bibinfo {volume} {111}},\ \bibinfo {pages} {145901} (\bibinfo {year}
  {2013})}\BibitemShut {NoStop}%
\bibitem [{\citenamefont {Sousa}\ \emph {et~al.}(2012)\citenamefont {Sousa},
  \citenamefont {Ferreira},\ and\ \citenamefont {Barroso}}]{SousaJCP2012}%
  \BibitemOpen
  \bibfield  {author} {\bibinfo {author} {\bibfnamefont {J.~M.~G.}\
  \bibnamefont {Sousa}}, \bibinfo {author} {\bibfnamefont {A.~L.}\ \bibnamefont
  {Ferreira}}, \ and\ \bibinfo {author} {\bibfnamefont {M.~A.}\ \bibnamefont
  {Barroso}},\ }\bibfield  {title} {\enquote {\bibinfo {title} {Determination
  of the solid-fluid coexistence of the n - 6 {L}ennard-{J}ones system from
  free energy calculations},}\ }\href {\doibase 10.1063/1.4707746} {\bibfield
  {journal} {\bibinfo  {journal} {J. Chem. Phys.}\ }\textbf {\bibinfo {volume}
  {136}},\ \bibinfo {pages} {174502} (\bibinfo {year} {2012})}\BibitemShut
  {NoStop}%
\bibitem [{\citenamefont {Thol}\ \emph {et~al.}(2016)\citenamefont {Thol},
  \citenamefont {Rutkai}, \citenamefont {K\"{o}ster}, \citenamefont {Lustig},
  \citenamefont {Span},\ and\ \citenamefont {Vrabec}}]{Thol2016}%
  \BibitemOpen
  \bibfield  {author} {\bibinfo {author} {\bibfnamefont {M.}~\bibnamefont
  {Thol}}, \bibinfo {author} {\bibfnamefont {G.}~\bibnamefont {Rutkai}},
  \bibinfo {author} {\bibfnamefont {A.}~\bibnamefont {K\"{o}ster}}, \bibinfo
  {author} {\bibfnamefont {R.}~\bibnamefont {Lustig}}, \bibinfo {author}
  {\bibfnamefont {R.}~\bibnamefont {Span}}, \ and\ \bibinfo {author}
  {\bibfnamefont {J.}~\bibnamefont {Vrabec}},\ }\bibfield  {title} {\enquote
  {\bibinfo {title} {Equation of state for the {L}ennard-{J}ones fluid},}\
  }\href {\doibase 10.1063/1.4945000} {\bibfield  {journal} {\bibinfo
  {journal} {J. Phys. Chem. Ref. Data}\ }\textbf {\bibinfo {volume} {45}},\
  \bibinfo {pages} {023101} (\bibinfo {year} {2016})}\BibitemShut {NoStop}%
\bibitem [{\citenamefont {Meier}(2002)}]{Meier2002}%
  \BibitemOpen
  \bibfield  {author} {\bibinfo {author} {\bibfnamefont {K.}~\bibnamefont
  {Meier}},\ }\href@noop {} {\emph {\bibinfo {title} {Computer Simulation and
  Interpretation of the Transport Coefficients of the {L}ennard-{J}ones Model
  Fluid (PhD Thesis)}}}\ (\bibinfo  {publisher} {Shaker},\ \bibinfo {address}
  {Aachen},\ \bibinfo {year} {2002})\BibitemShut {NoStop}%
\end{thebibliography}%

\end{document}